\documentclass[a4paper,11pt]{article}
\pdfoutput=1 
\usepackage{jheppub}
\usepackage[T1]{fontenc} 
\usepackage{color}
\usepackage{fancybox}
\usepackage{amsmath}
\usepackage{amsfonts}
\usepackage{slashed}
\usepackage{amssymb}

\def\Tr{{\rm Tr\, }}

\newcommand{\be}{\begin{equation}}
\newcommand{\bea}{\begin{eqnarray}}

\newcommand{\ee}{\end{equation}}
\newcommand{\eea}{\end{eqnarray}}

\def\a{\alpha}\def\b{\beta}

\title{\boldmath COMMENTS ON THE T-DUAL OF THE GRAVITY DUAL OF D5-BRANES ON $S^3$}

\author{THIAGO R. ARAUJO}
\author{and HORATIU NASTASE}

\affiliation{Instituto de F\'{i}sica Te\'{o}rica, UNESP-Universidade Estadual Paulista\\ R. Dr. Bento T. Ferraz 271, Bl. II, Sao Paulo 01140-070, SP, Brazil}

\emailAdd{taraujo@ift.unesp.br}
\emailAdd{nastase@ift.unesp.br}

\abstract{We consider an abelian T-duality on a deformation of the gravitational solution of \cite{Maldacena2001}, which is the gravity dual of 
$N_c$ D5-branes wrapping a three-cycle inside a manifold that admits a $G_2$ structure. Performing the T-duality we find $N_c$ D$4$-branes 
wrapping a two-cycle with non-trivial antisymmetric fields in the NS-NS and RR sector. We study some aspects of its dual field theory and 
we compare with the original solution.}


\begin{document} 

\maketitle

\section{Introduction}

Gauge/gravity correspondence has been used to explore many aspects of gauge theories which cannot be studied using usual perturbation theory techniques.
The original correspondence was proposed as a duality between the $AdS_5\times S^5$ background of type IIB supergravity 
and the $\mathcal{N}=4$ SYM in four dimensions \cite{Maldacena1997, Aharony1999a}.

One important feature of this original correspondence is that the duality relates string theory and a conformal field theory with maximal supersymmetry 
and all fields transforming in the adjoint representation. To make contact with the real world, we need to extend these ideas to non-conformal 
field theories with minimal supersymmetry as well as adding fields transforming in the fundamental representation.

In particular, for phenomenological applications we need ${\cal N}=1$ supersymmetry. In \cite{Maldacena2000b}, the authors found the gravity dual 
of a pure $\mathcal{N}=1$ SYM in $d=3+1$ (coupled to extra modes that could not be decoupled while maintaining calculability). In this particular solution, 
called the Maldacena-N\'{u}\~{n}ez solution,  we start with $N_c$ D$5$-branes, where the field theory living on the worldvolume of these branes carries $16$ supercharges, and we wrap them on a sphere $S^2$. In general this breaks supersymmetry. In order to preserve some fraction of the 
original supersymmetries, we twist the fields in such a way that we preserve four supercharges \cite{Bershadsky1995, Maldacena2000}, 
equivalent to $\mathcal{N}=1$ supersymmetry in $3+1$ dimensions.

As we already mentioned, realistic theories require fields transforming in the fundamental representation. To address this, one considers 
flavor branes in the gravity side, which is equivalent to adding an open string sector \cite{Karch2002}. One can start by studying 
the {\it quenched} approximation, when probe branes are used in a way that the number of flavor branes $N_f$ in negligible compared to the number 
$N_c$ of color branes. Then, the next natural step is to consider the unquenched case, that is the case in which the number 
of flavor branes is of the same order as the number color branes \cite{Nunez2003, Nunez2006, Nunez2010}.

Another important development was considered in \cite{Chamseddine2001, Schvellinger2001, Maldacena2001},  where it was found a solution of 5
dimensional supergravity which can be lifted to 7 dimensions and then to 10 dimensions. In this case we have a gravitational solution that 
holographically describes D5-branes wrapping a three-cycle inside a $G_2$ manifold. In the IR limit, the theory living in the worldvolume of 
these branes was identified as being dual to $\mathcal{N}=1$ $SU(N_c)$ SYM in three-dimensions with Chern-Simons level $k=N_c /2$. 

In \cite{Canoura2008} the ansatz of \cite{Chamseddine2001} was generalized and this allowed one to find a new class of solutions in which in the 
UV limit the metric is a product of a $G_2$ cone and a three dimensional Minkowski space, and the dilaton is a constant, in contrast to the original 
behaviour of the Maldacena-Nastase solution, where the dilaton diverges as the holographic coordinate goes to infinity. It is important to realize that 
this solution corresponds to D$5$-branes wrapped on a three-cycle of a $G_2$ cone in which the near-horizon effects of the branes on the metric become 
negligible in the UV limit. 

Note that Canoura, Merlatti and Ramallo \cite{Canoura2008} also added massless fundamental flavors to the Maldacena-Nastase (hereafter MNa) solution 
in the unquenched case. The authors found that this system with $N_f\geq2 N_c$ dramatically differs from $N_f<2 N_c$. Massive fundamental 
flavors were added to the MNa solution in \cite{Macpherson2014} and the author showed that is is possible to find a solution which interpolates 
between the deformed unflavored MNa background and the massless flavored background. 

As pointed in \cite{Macpherson2013b, Macpherson2013}, we can obtain the UV completion of this solution considering a $G_2$-structure 
rotation \cite{Gaillard2011} which is a solution generating technique analogous to the U-duality. The rotation procedure is implemented in a type 
IIA with $\mathcal{N}=1$ and generates a more general type IIA solution. The important point is that in this rotation procedure we have an extra 
warp factor in the metric and this term ensures the finiteness of the cycle along the energy scale.

The gauge theory analysis of the rotated MNa solution was performed in \cite{Macpherson2013b}, and the author showed that the dual field theory is 
confining and that in the IR limit the Chern-Simons term dominates the dynamics of the theory.

Another well-known important generating solution technique is T-duality. In \cite{Macpherson2013} a non-abelian T-duality has been considered 
along the $SU(2)$ isometry of the deformed MNa solution \cite{Canoura2008}, and this gave a rather complicated massive type IIA solution, with 
all fields in the RR sector and which is dual to a confining Chern-Simons gauge theory.

In this article we consider a T-duality transformation on the MNa solution. In section 2 we start with a review of the solution due to 
Canoura et al. \cite{Canoura2008}, and which contains the original solution \citep{Maldacena2001} as a special case. Next, 
in section 3 we start to perform an 
abelian T-duality in the MNa solution along an $U(1)$ isometry in the D$5$-brane solution, which gives a D$4$-brane solution wrapping a two-cycle. 
We close section 3 by computing the Maxwell and Page charges.

In section 4 we consider some aspects of the dual gauge theory. In section 4.1 we find the quark-antiquark potential and we see that the requirements for confinement are satisfied. In such a case we are able to compute the string tension. Next we follow considering the gauge coupling and the entanglement entropy, which has been used as a probe of confinement. Finally, we study some conditions in which we can treat the wrapped D$4$-branes as a domain wall, so that we induce a Chern-Simons term in the gauge theory.

\section{Wrapped fivebranes on a three-cycle} \label{mald-nast}

We start our analysis by considering a short review of the deformed MNa solution \cite{Maldacena2001}. It is a type-IIB supergravity solution that consists of D$5$-branes wrapping  a $3$-cycle in a manifold that supports a $G_2$-structure. In the IR limit this theory is dual to $\mathcal{N}=1$ SYM in three 
dimensions. In \cite{Canoura2008} the ansatz was generalized and this solution has the original solution as a special case. The string frame metric 
is given by
\begin{equation}
 ds^{2}_{st}=e^\varphi \left(dx^{2}_{1,2}+ds_{7}^2\right),
\end{equation}
and the internal part of the metric, which describes the manifold supporting a $G_2$-structure, is
\begin{equation}
 ds_{7}^{2}=N_c\left[e^{2g}dr^2+\frac{e^{2h}}{4}(\sigma^i)^2+\frac{e^{2g}}{4}\left(\omega^i-\frac{1}{2}(1+w)\sigma^i\right)^2\right],
\end{equation}
where we are using an optimum holographic coordinate defined in \cite{Macpherson2013b}. Also, $\sigma^i$ and $\omega^i$ are two sets of $SU(2)$ Maurer-Cartan forms satisfying
\begin{equation}
 d\lambda_{a}^{i}=-\frac{1}{2}\epsilon_{ijk}\lambda_{a}^{j}\wedge \lambda_{a}^{k},
\end{equation}
where $\lambda_{1}^{i}=\sigma^{i}$ and $\lambda_{2}^{i}=\omega^{i}$ for $i=1,2,3$. These forms can be represented in terms of Euler angles as
\begin{subequations}
\begin{align}
 \lambda_{a}^{1}&= \cos \psi_a d\theta_a+\sin \psi_a\sin \theta_a d\phi_a\\
 \lambda_{a}^{2}&= - \sin \psi_a d\theta_a+\cos \psi_a\sin \theta_a d\phi_a\\
 \lambda_{a}^{3}&= d\psi_a+\cos \theta_a d\phi_a\;,
\end{align}
\end{subequations}
for $0\leq \theta_a \leq \pi$, \ $0\leq \phi_a < 2\pi$, \ $0\leq \psi_a < 4\pi$. 

Also, the MNa solution has a non-trivial RR $3$-form
\begin{equation}
\begin{split}
F_3 &= \frac{N_c}{4}\left\lbrace (\sigma^1\wedge\sigma^2\wedge\sigma^3-\omega^1\wedge\omega^2\wedge\omega^3)+\frac{\gamma^\prime}{2}dr\wedge \sigma^i\wedge \omega^i-\right.\\
&\hspace{1.5cm}-\left.\frac{(1+\gamma)}{4}\epsilon_{ijk }[\sigma^i\wedge\sigma^j\wedge \omega^k-\omega^i\wedge\omega^j\wedge \sigma^k] \right\rbrace.
\end{split}\label{threeform}
\end{equation}
One can easily show that this field strength is generated by the following two-form potential
\begin{align}
\label{potent}
 C^{(2)} &= \left(-\frac{1}{4}N_c \cos\theta_1\right)d\phi_1\wedge d\psi_1+\left(\frac{1}{4}N_c \cos\theta_2\right)d\phi_2\wedge d\psi_2+\frac{N_c(1+\gamma)}{8}d \psi_1\wedge d\psi_2+\nonumber\\
 & +\frac{N_c(1+\gamma)}{8}\sin\theta_1\sin(\psi_1-\psi_2)d\phi_1\wedge d\theta_2+\nonumber\\
 &+\frac{N_c(1+\gamma)}{8}(\cos\theta_1\cos\theta_2+\sin\theta_1\sin\theta_2\cos(\psi_1-\psi_2))d\phi_1\wedge d\phi_2\\
 &+\frac{N_c(1+\gamma)}{8}\cos\theta_1 d\phi_1\wedge d\psi_2+\frac{N_c(1+\gamma)}{8}\cos (\psi_1-\psi_2)d\theta_1\wedge d\theta_2\nonumber\\
 &+\left(-\frac{N_c(1+\gamma)}{8}\sin(\psi_1-\psi_2)\sin\theta_2\right)d\theta_1\wedge d\phi_2+\frac{N_c(1+\gamma)}{8}\cos\theta_2 d \psi_1\wedge d\phi_2\;,
\nonumber
\end{align}
so that $F_3=dC^{(2)}$.

Unfortunately, the solution for these equations is known just semi-analytically in the IR and UV limits. In the IR limit, that is, $r\sim 0$ we have
\begin{subequations}
\begin{align}
e^{2g}&=g_0+\frac{(g_0-1)(9g_0+5)}{12g_0}r^2+\dots\label{iri}\\
e^{2h}&=g_0 r^2-\frac{3g_0^2-4g_0+4)}{18g_0}r^4+\dots\\
w&=1-\frac{3g_0-2}{3g_0}r^2+\dots\\
\gamma &=1-\frac{1}{3}r^2+\dots \\
\phi&=\phi_0+\frac{7}{24 g_0^2}r^2.\label{irf}
\end{align}
\end{subequations}

On the other hand, in the UV limit, where $r\sim\infty$, we have
\begin{subequations}
\begin{align}
e^{2g}&=c_1 e^{4r/3}-1+\frac{33}{4c_1}e^{-4r/3}\label{uvi}\\
e^{2h}&=\frac{3c_1}{4} e^{4r/3}+\frac{9}{4}-\frac{77}{16c_1}e^{-4r/3}\\
w&=\frac{2}{c_1}e^{-4r/3}+\dots\\
\gamma &=\frac{1}{3}+\dots\\
\phi&=\phi_\infty+\frac{2}{c_1^2}e^{-8r/3}.\label{uvf}
\end{align}
\end{subequations}

We write the whole set of components of the string frame metric as $$x^M=\{x^\mu, x^A\};\quad \{(\mu=0,1,2);(A=r,\tilde{\alpha}, \alpha)\},$$ where
$$\{x^r\equiv r; x^{\tilde{\alpha}}\equiv\theta_1,\phi_1,\psi_1;x^\alpha=\theta_2,\phi_2,\psi_2\}. $$
Now we have
\begin{equation}
 (\lambda_{a}^{i})^2=d\theta_{a}^{2}+d\phi_{a}^{2}+d\psi_{a}^{2}+2\cos\theta_a d\psi_a d\phi_a \label{lambdasqr}
 \end{equation}
 and
 \begin{align}
\omega^{i}\sigma^{i}&=\cos(\psi_1-\psi_2)d\theta_1 d\theta_2-\sin(\psi_1-\psi_2)\sin\theta_2 d\theta_1 d\phi_2\nonumber\\
&+\sin(\psi_1-\psi_2)\sin\theta_1 d\phi_1 d\theta_2+[\cos(\psi_1-\psi_2)\sin\theta_1\sin\theta_2+\cos\theta_1\cos\theta_2]d\phi_1 d\phi_2 \label{omegasigma}\\
&+\cos\theta_1 d\phi_1 d\psi_2+\cos\theta_2 d\psi_1d\phi_2+d\psi_1 d\psi_2\nonumber.
\end{align}
Then, we write the string-frame metric as
\begin{equation}
 ds_{st}^{2}=g_{MN}dx^M dx^N=e^\varphi dx_{1,2}^2+\Delta dr^2+\Sigma (\sigma^i)^2+\Omega(\omega^i)^2+2\Xi\omega^i\sigma^i\;,
\end{equation}
where we define
\begin{subequations}
\begin{align}
 \Delta &=e^{\varphi+2g}N_c\\
 \Sigma&=\frac{e^{\varphi}}{4}N_c\left(e^{2h}+\frac{e^{2g}}{4}(1+w)^2\right)\equiv e^\varphi \widetilde{\Sigma}\\
 \Omega&=\frac{e^{\varphi+2g}}{4}N_c\equiv \frac{\Delta}{4}\\
 \Xi&=-\frac{e^{\varphi+2g}}{8}(1+w)N_c\equiv  - \frac{\Omega}{2}(1+w)
\end{align}
\end{subequations}
for later convenience. Finally, using that $M=\{\mu, A\}$ we find the components of the metric matrix 
$$
(g_{MN})=
\begin{pmatrix}
  g_{\mu\nu}=e^\varphi \eta_{\mu\nu}\quad & g_{\mu A}=0 \\
  g_{A \mu}=0\quad & g_{AB}
 \end{pmatrix}.
$$
Obviously, we need to find just the components $g_{AB}$ and these are
$$
\begin{array}{|l|l|l|}
  \hline
  g_{rr}=\Delta & g_{r\mu}=0 & g_{r\tilde{\alpha}}=g_{r\alpha}=0 \\
  \hline
 \end{array}
$$
$$
\begin{array}{|l|l|}
  \hline
  \ g_{\tilde{\theta}\tilde{\theta}}=g_{\tilde{\phi}\tilde{\phi}}=g_{\tilde{\psi}\tilde{\psi}}=\Sigma\ &\ g_{\tilde{\phi}\tilde{\psi}}=\Sigma\cos\theta_1 \\
  \hline
  \ g_{\theta\theta}=g_{\phi\phi}=g_{\psi\psi}=\Omega\ &\ g_{\phi\psi}=\Omega\cos\theta_2 \\
  \hline
 \end{array}
$$
$$
\begin{array}{|l|l|l|}
  \hline
  g_{\theta \tilde{\theta}}=\Xi\cos(\psi_1-\psi_2) & g_{\phi \tilde{\theta}}=-\Xi\sin(\psi_1-\psi_2) \sin\theta_2 & g_{\psi \tilde{\theta}}=0 \\
  \hline
  g_{\theta \tilde{\phi}}=\Xi\sin(\psi_1-\psi_2)\sin\theta_1 & g_{\phi \tilde{\phi}}=\Xi[\sin\theta_1\sin\theta_2\cos(\psi_1-\psi_2)+ \cos\theta_1\cos\theta_2] & g_{\psi \tilde{\phi}}=\Xi\cos\theta_1 \\
  \hline
  g_{\theta \tilde{\psi}}=0 & g_{\phi \tilde{\psi}}=\Xi\cos\theta_2 & g_{\psi \tilde{\psi}}=\Xi \\
  \hline
 \end{array}
$$

\section{D4-brane solution}

Now we perform a T-duality transformation in a direction along the brane, namely, the $x^{\tilde{\phi}}\equiv\phi_1$ direction. If we consider the type-IIA solution with NS-NS sector given by $\{\tilde{\varphi},\ \tilde{g}_{MN},\ \mathcal{B}_{MN}\}$, the Buscher's rules \cite{Buscher1988, Ortin2004, Bergshoeff1995, Johnson2003} are
\begin{center}
\shadowbox{\begin{minipage}{3.8in}
$$e^{2\tilde{\varphi}}=\frac{e^{2\varphi}}{|g_{\tilde{\phi}\tilde{\phi}}|} \hspace{1.5cm} \tilde{g}_{\tilde{\phi}\tilde{\phi}}=\frac{1}{g_{\tilde{\phi}\tilde{\phi}}} $$
$$\tilde{g}_{MN}=g_{MN}-\frac{g_{\tilde{\phi} M} g_{\tilde{\phi} N}-B_{\tilde{\phi} M} B_{\tilde{\phi} N}}{g_{\tilde{\phi}\tilde{\phi}}}\qquad \tilde{g}_{\tilde{\phi}M}=\frac{1}{g_{\tilde{\phi}\tilde{\phi}}}B_{\tilde{\phi}M}  $$
$$\mathcal{B}_{MN}=B_{MN}-2\frac{B_{\tilde{\phi} [M} g_{ N]\tilde{\phi}}}{g_{\tilde{\phi}\tilde{\phi}}}\qquad \mathcal{B}_{M \tilde{\phi} }=-\frac{g_{M \tilde{\phi} }}{g_{\tilde{\phi}\tilde{\phi}}}  $$
\end{minipage}}
\end{center}

\subsection{NS-NS sector} \label{nsnssec1}

Using the transformation rules above, the dilaton is
\begin{equation}
 e^{2\tilde{\varphi}}=\frac{1}{\Sigma}e^{2\varphi}=\frac{1}{\widetilde{\Sigma}}e^{\varphi}
\end{equation}
and the dual metric is
\begin{align}
 d\tilde{s}_{st}^2=&e^{2\tilde{\varphi}}\widetilde{\Sigma}dx_{1,2}^2+\Delta dr^2+\frac{1}{\Sigma}d\phi_1^2+\Sigma (d\theta_1^2+\sin^2\theta_1 d\psi_1^2)\nonumber\\
 &+2\Xi\left[(\cos\psi_1\omega^{1}-\sin\psi_1\omega^{2})d\theta_1-
 \sin\theta_1\cos\theta_1(\sin\psi_1\omega^{1}+\cos\psi_1\omega^{2})d\psi_1\right.\nonumber\\
 &+\left.\sin^2\theta_1\omega^{3}d\psi_1\right]+\Omega (\omega^i)^2\label{dualmetric}\\
 &-\frac{\Xi^2}{\Sigma}\left[\sin^2\psi_1\sin^2\theta_1(\omega^1)^2+2\sin\psi_1\cos\psi_1\sin^2\theta_1\omega^1\omega^2\right.\nonumber\\
 &+2\sin\psi_1\cos\theta_1\sin\theta_1\omega^1\omega^3+\cos^2\psi_1\sin^2\theta_1(\omega^2)^2+2\cos\psi_1\sin\theta_1\cos\theta_1\omega^2\omega^3\nonumber\\
 &+\left.\cos^2\theta_1(\omega^3)^2\right]\;,\nonumber
\end{align}
where we can rewrite the coefficients in terms of the type-IIA dilaton $\tilde{\varphi}$
\begin{align}
 \Delta &=e^{2\tilde{\varphi}+2g}N_c\widetilde{\Sigma}\cr
 \Sigma&=e^{2\tilde{\varphi}} \widetilde{\Sigma}^2\cr
 \Omega&=\frac{e^{2\tilde{\varphi}+2g}}{4}N_c\widetilde{\Sigma}\cr
 \Xi&=-\frac{e^{2\tilde{\varphi}+2g}}{8}(1+w)N_c\widetilde{\Sigma}.
\end{align}
Also, we define a first rotation
\begin{align}
 \tilde{\omega}^1&=\cos\psi_1\omega^1-\sin\psi_1\omega^2=\cos(\psi_2-\psi_1)d\theta_2+\sin(\psi_2-\psi_1)\sin\theta_2d\phi_2\cr
 \tilde{\omega}^2&=\sin\psi_1\omega^1+\cos\psi_1\omega^2=-\sin(\psi_2-\psi_1)d\theta_2+\cos(\psi_2-\psi_1)\sin\theta_2d\phi_2\cr
 \tilde{\omega}^3&=\omega^3=d\psi_2+\cos\theta_2d\phi_2\cr
 \tilde{\sigma}^1&=\cos\psi_1\sigma^1-\sin\psi_1\sigma^2\cr
 \tilde{\sigma}^2&=\sin\psi_1\sigma^1+\cos\psi_1\sigma^2\cr
 \tilde{\sigma}^3&=\sigma^3\;.
\end{align}
We then consider a second rotation
\begin{align}
 \hat{\omega}^1&=\tilde{\omega}^1\cr
 \hat{\omega}^2&=\cos\theta_1\tilde{\omega}^2-\sin\theta_1\tilde{\omega}^3\cr
 \hat{\omega}^3&=\sin\theta_1\tilde{\omega}^2+\cos\theta_1\tilde{\omega}^3\cr
 \hat{\sigma}^1&=\tilde{\sigma}^1\cr
 \hat{\sigma}^2&=\cos\theta_1\tilde{\sigma}^2-\sin\theta_1\tilde{\sigma}^3\cr
 \hat{\sigma}^3&=\sin\theta_1\tilde{\sigma}^2+\cos\theta_1\tilde{\sigma}^3\;,
\end{align}
obtaining the metric
\begin{align}
 d\tilde{s}_{st}^2=&\frac{N_c}{4}e^{2\tilde{\varphi}}\left(e^{2h}+\frac{e^{2g}}{4}(1+w)^2\right)dx_{1,2}^2+\Delta dr^2+\frac{1}{\Sigma}d\phi_1^2+\Sigma (d\theta_1^2+\sin^2\theta_1 d\psi_1^2)\cr
 &+2\Xi[\tilde{\omega}^{1}d\theta_1-\sin\theta_1\cos\theta_1\tilde{\omega}^{2}d\psi_1+\sin^2\theta_1\tilde{\omega}^{3}d\psi_1]+\Omega (\tilde{\omega}^{i})^2\cr
 &-\frac{\Xi^2}{\Sigma}[\sin\theta_1\tilde{\omega}^2+\cos\theta_1\tilde{\omega}^3]^2\;,
 \end{align}
or reorganizing
 \begin{align}
 d\tilde{s}_{st}^2=&\frac{N_c}{4}e^{2\tilde{\varphi}}\left(e^{2h}+\frac{e^{2g}}{4}(1+w)^2\right)dx_{1,2}^2+\Delta dr^2+\frac{1}{\Sigma}d\phi_1^2+\nonumber\\
 &+\left(\Sigma-e^{2\tilde{\varphi}+2g}\frac{(1+w)^2}{4^2}N_c\widetilde{\Sigma}\right) (d\theta_1^2+\sin^2\theta_1 d\psi_1^2)\nonumber\\
 &+e^{2\tilde{\varphi}+2g}\frac{N_c}{4}\widetilde{\Sigma}\left[\left(\hat{\omega}^1-\frac{1}{2}(1+w)d\theta_1\right)^2+
 \left(\hat{\omega}^2+\frac{1}{2}(1+w)\sin\theta_1 d\psi_1\right)^2\right]\nonumber\\
 &+\left(\Omega-\frac{\Xi^2}{\Sigma}\right)(\hat{\omega}^3)^2.
\end{align}

Finally, the $2$-form field, which vanishes in the original solution, is non-trivial after the T-duality and one can write in the following form 
\begin{equation}
\begin{split}
 \mathcal{B}=&-\left\{\cos\theta_1 d\psi_1\wedge d\phi_1+\frac{\Xi}{\Sigma} \sin(\psi_1-\psi_2)\sin\theta_1 d\theta_2\wedge d\phi_1\right.\\
 &\left.+\frac{\Xi}{\Sigma} [\sin\theta_1\sin\theta_2\cos(\psi_1-\psi_2)+\cos\theta_1\cos\theta_2]d\phi_2\wedge d\phi_1+\frac{\Xi}{\Sigma} \cos\theta_1 d\psi_2\wedge d\phi_1\right\}.
\end{split}
\end{equation}

One important cycle in this background is 
\begin{equation}
\mathcal{C}_2=\lbrace\theta_1=\theta_2\equiv \theta;\psi_1=\psi_2\equiv \psi|\phi_1,\phi_2,r,x_1,x_2={\rm const.}\rbrace
\end{equation}
which is the cycle where the metric is wrapped. The induced metric is given by
\begin{equation}
ds_{\mathcal{C}_2}^2=\left(\Sigma+2\Xi+\Omega\right)d\theta^2+\left(\Omega+\Sigma\sin^2\theta+2\Xi\sin^2\theta-\frac{\Xi^2}{\Sigma}\cos^2\theta\right)d\psi^2\;,
\end{equation}
and vanishes in the IR limit. The B field vanishes on this cycle.

\subsection{R-R sector}

Remember that the RR-sector for the type IIA supergravity is $\{C^{(1)}, C^{(3)}\}$ while the RR-sector for type IIB supergravity is $\{C^{(0)}, C^{(2)}, C^{(4)}\}$ and in the present case, the non-trivial field is just $C^{(2)}$, whose field strength is given by the $F_3$ in (\ref{threeform}), 
\begin{align}
 F_3 &= \frac{N_c}{4}\left\lbrace (\sigma^1\wedge\sigma^2\wedge\sigma^3-\omega^1\wedge\omega^2\wedge\omega^3)+\frac{\gamma^\prime}{2}dr\wedge \sigma^i\wedge \omega^i-\right.\nonumber\\
  &\hspace{1.5cm}-\left.\frac{(1+\gamma)}{4}\epsilon_{ijk }[\sigma^i\wedge\sigma^j\wedge \omega^k-\omega^i\wedge\omega^j\wedge \sigma^k].\right\rbrace
\end{align}

Given the T-duality rules for going from type-IIB to type-IIA supergravity,
\begin{subequations}
\begin{equation}
\begin{split}
C^{(2n+1)}_{M_1\dots M_{2n+1}}&=C^{(2n+2)}_{M_1\dots M_{2n+1}\tilde{\phi}}+(2n+1)B_{[M_1|\tilde{\phi}|}C^{(2n)}_{M_2\dots M_{2n+1}]}\\
&+2n(2n+1)B_{[M_1|\tilde{\phi}|}g_{M_2|\tilde{\phi}|}C^{(2n)}_{M_3\dots M_{2n+1}]\tilde{\phi}}/g_{\tilde{\phi}\tilde{\phi}}
\end{split}
\end{equation}
\begin{equation}
C^{(2n+1)}_{M_1\dots M_{2n}\tilde{\phi}}=C^{(2n)}_{M_1\dots M_{2n}}-2n g_{[M_1|\tilde{\phi}|}C^{(2n)}_{M_2\dots M_{2n}]\tilde{\phi}}/g_{\tilde{\phi}\tilde{\phi}}
\;, 
\end{equation}
\end{subequations}
we can use (\ref{potent}) and find the $RR$ potential forms of the type IIA-solution
\paragraph{n=0}\ $\phantom{x}$ \\
In this case, we have the following components of the dual theory
\begin{align}
C^{(1)}_{M_1}&=C_{M_1\tilde{\phi}}^{(2)}\cr
C^{(1)}_{\tilde{\phi}}&=C^{(0)}=0\;,
\end{align}
so we obtain the potential
\begin{equation}
\begin{split}
C^{(1)}=&-\left\{\frac{N_c(1+\gamma)}{8}[\cos\theta_1\cos\theta_2+\sin\theta_1\sin\theta_2\cos(\psi_1-\psi_2)]d\phi_2-\frac{N_c}{4}\cos\theta_1 d\psi_1\right.\\
&\left.+\frac{N_c(1+\gamma)}{8}\sin(\psi_1-\psi_2)\sin\theta_1 d\theta_2+\frac{N_c}{8}(1+\gamma)\cos\theta_1 d\psi_2\right\}.
\end{split}
\end{equation}

\paragraph{n=1}\ $\phantom{x}$ \\
In this case, we have
\begin{align}
C^{(3)}_{M_1 M_2 M_3}&=C_{M_1 M_2 M_3\tilde{\phi}}^{(4)}=0\cr
C^{(3)}_{M_1 M_2 \tilde{\phi}}&=C^{(2)}_{M_1 M_2}-\frac{1}{g_{\tilde{\phi}\tilde{\phi}}}(g_{M_1 \tilde{\phi}}C_{M_2 \tilde{\phi}}^{(2)}-g_{M_2 \tilde{\phi}}C_{M_1 \tilde{\phi}}^{(2)}).
\end{align}
Therefore we obtain the three-form potential
\begin{align}
C^{(3)}&=-\frac{N_c(1+\gamma)}{8}\cos(\psi_1-\psi_2)d\theta_1\wedge d\phi_1\wedge d\theta_2\nonumber\\
&+\frac{N_c(1+\gamma)}{8}\sin\theta_2\sin(\psi_1-\psi_2)d\theta_1\wedge d\phi_1\wedge d\phi_2\nonumber\\
&-\frac{N_c}{8\Sigma}\cos\theta_1\sin\theta_1\sin(\psi_1-\psi_2)\left[2\Xi+\Sigma(1+\gamma)\right]d\psi_1\wedge d\phi_1\wedge d\theta_2\nonumber\\
&-\left[\frac{N_c(1+\gamma)}{8}\cos\theta_2\right. \\
&\left.+\frac{N_c}{4\Sigma}\left(\Xi+\frac{(1+\gamma)}{2}\Sigma\right)\cos\theta_1[\cos\theta_1\cos\theta_2+
\sin\theta_1\sin\theta_2\cos(\psi_1-\psi_2)]\right]d\psi_1\wedge d\phi_1\wedge d\phi_2\nonumber\\
&-\left[\frac{N_c(1+\gamma)}{8}+\frac{N_c}{4\Sigma}\left(\Xi+\frac{\Sigma}{2}(1+\gamma)\right)\cos^2\theta_1\right]d\psi_1\wedge d\phi_1\wedge d\psi_2\nonumber\\
&+\frac{N_c}{4}\cos\theta_2 d\phi_1\wedge d\phi_2\wedge d\psi_2.\nonumber
\end{align}

We have generated a type IIA-solution of supergravity wich consists of $N_c$ D$4$-branes wrapping a two-cycle and with a perpendicular $S^1$ manifold. 
This solution has non-trivial RR $2$ and $4$-forms defined by $F_2=dC^{(1)}$ and $F_4=dC^{(3)}$. 

For completeness, starting from a solution of supergravity in eleven dimensions, one can consider a dimensional reduction on a circle $S^1$ to 
a type-IIA solution. Conversely, given a  solution of the type-IIA supergravity, we can lift it to a solution of eleven dimensional supergravity. In fact,
the eleven dimensional fields corresponding to the type IIA ones are written as 
\begin{center}
\shadowbox{\begin{minipage}{4.5in}
$$g^{(11)}_{MN}=e^{-2\tilde{\varphi}/3}\tilde{g}_{MN}+e^{4\tilde{\varphi}/3}C^{(1)}_{M}C^{(1)}_{N}\qquad (C^{(3)})^{11}_{MNP}=C^{(3)}_{MNP} $$
$$g^{(11)}_{M,11}=e^{4\tilde{\varphi}/3}C^{(1)}_{M}\qquad (C^{(3)})^{11}_{MN,11}= \mathcal{B}_{MN} $$
$$g^{(11)}_{11,11}=e^{4\tilde{\varphi}/3}$$
\end{minipage}}
\end{center}
Rewriting the dual metric (\ref{dualmetric}) as
\begin{align}
 d&\tilde{s}_{st}^2=\frac{N_c}{4}e^{2(\tilde{\varphi})}\left(e^{2h}+\frac{e^{2g}}{4}(1+w)^2\right)dx_{1,2}^2+\Delta dr^2+\frac{1}{\Sigma}d\phi_1^2+\Sigma (d\theta_1^2+\sin^2\theta_1 d\psi_1^2)\nonumber\\
 &+2\Xi[\cos(\psi_1-\psi_2)d\theta_1 d\theta_2-\sin(\psi_1-\psi_2)\sin\theta_2 d\theta_1 d\phi_2-\sin(\psi_1-\psi_2)\sin\theta_1\cos\theta_1 d\psi_1 d\theta_2\nonumber\\
 &(\cos\theta_2\sin^2\theta_1-\cos\theta_1\sin\theta_1\sin\theta_2\cos(\psi_1-\psi_2))d\psi_1 d\phi_2+\sin^2\theta_1 d\psi_1 d\psi_2]\nonumber\\
 &+\left(\Omega-\frac{\Xi^2}{\Sigma}\sin^2(\psi_1-\psi_2)\sin^2\theta_1\right)d\theta_{2}^{2}\nonumber\\
 &+\left(\Omega-\frac{\Xi^2}{\Sigma}[\sin\theta_1\sin\theta_2\cos(\psi_1-\psi_2)+\cos\theta_1\cos\theta_2]^2\right)d\phi_{2}^{2}\nonumber\\
 &+\left(\Omega-\frac{\Xi^2}{\Sigma}\cos^2\theta_1\right)d\psi_{2}^{2}+\nonumber\\
 &+2\left(\Omega\cos\theta_2-\frac{\Xi^2}{\Sigma}\cos\theta_1[\sin\theta_1\sin\theta_2\cos(\psi_1-\psi_2)+\cos\theta_1\cos\theta_2]\right)  d\phi_2 d\psi_2\nonumber\\
 &-2\frac{\Xi^2}{\Sigma}[\sin\theta_1\sin\theta_2\cos(\psi_1-\psi_2)+\cos\theta_1\cos\theta_2]\sin(\psi_1-\psi_2)\sin\theta_1 d\theta_2 d\phi_2\nonumber\\
 &-2\frac{\Xi^2}{\Sigma}\sin(\psi_1-\psi_2)\sin\theta_1 \cos\theta_1 d\theta_2 d\psi_2\;,
\end{align}
the eleven dimensional metric becomes
\begin{align}
 ds_{(11)}^2&=e^{-2\tilde{\varphi}/3}d\tilde{s}_{st}^2\cr &+e^{4\tilde{\varphi}/3}\left(C^{(1)}_{\tilde{\psi}}C^{(1)}_{\tilde{\psi}}+C^{(1)}_{\tilde{\psi}}C^{(1)}_\phi+C^{(1)}_{\tilde{\psi}}C^{(1)}_\psi+C^{(1)}_{\tilde{\psi}}C^{(1)}_\theta\right.\cr &\left.+C^{(1)}_{\phi}C^{(1)}_{\phi}+C^{(1)}_{\phi}C^{(1)}_\psi+C^{(1)}_{\phi}C^{(1)}_\theta+C^{(1)}_{\psi}C^{(1)}_\psi+C^{(1)}_{\psi}C^{(1)}_\theta+C^{(1)}_\theta C^{(1)}_\theta\right)\cr
&e^{4\tilde{\varphi}/3}\left(C^{(1)}_{\tilde{\psi}}+C^{(1)}_\phi+C^{(1)}_\psi+C^{(1)}_\theta+dx^{10}\right)dx^{10}.
\end{align}

\subsection{Brane charges}

Superstring theories have massless $p$-form potentials which may be regarded as generalizations of the electromagnetic gauge field. 
The Maxwell equations for the gauge field of electrodynamics $A^{(1)}=A_\mu dx^\mu$ in the presence of sources are
\begin{equation}
dF_2=\star \mathcal{J}_m\,,\qquad d\star F_2=\star \mathcal{J}_e.
\end{equation}
It follows that the electric and magnetic charges are given by
\begin{equation}
e=\int_{S^2}\star F_2\,,\qquad g=\int_{S^2} F_2\;,
\end{equation}
where $S^2$ is a two-sphere surrounding the charges.

In string theory in the presence of $n$-forms, we can define conserved charges associated to the gauge potentials and then find the stable branes of
given electric charge. 
For instance, a D$p$-brane in type II superstring theory couples to a $(p+1)$-form $C^{(p+1)}$ with field strength $F_{p+2}=dC^{(p+1)}$. 
The corresponding electric-type charge is
\begin{equation}
\mathcal{Q}_{Dp}=\int_{\Sigma^{8-p}}\star F_{p+2}\;,
\end{equation}
where $\mathcal{C}^{8-p}$ is a cycle surrounding the charge. 

As an explicit example, consider the original background reviewed in section \ref{mald-nast}. We know that this solution corresponds to $N_c$ D$5$branes
on an $S^3$. Consider then the 3-cycle
\begin{equation}
\tilde{S}^3=\lbrace \omega^i |\sigma^i=0\rbrace,
\end{equation}
and integrate the RR three form (\ref{threeform}) on it, obtaining ($\int \omega^1\wedge\omega^2\wedge\omega^3=16\pi^2$)
\begin{equation}
\frac{1}{4\pi^2}\int_{\tilde{S}^3} F_{3}=N_c,
\end{equation}
which means that we have a quantization condition.

In \cite{Marolf2000}, the author showed that there are different types of electric or magnetic charge associated with a gauge field. 
Here we collect the main results for D$4$-branes, which is the case we are interested in. 

In the T-dual solution that we computed above, we have one non trivial RR $1$-form $C^{(1)}$ and one $3$-form $C^{(3)}$, and the Kalb-Ramond field 
$\mathcal{B}$ is also non-vanishing. The $4$-form gauge field, which is invariant under the abelian gauge transformation $C^{(1)}\to C^{(1)}+d\xi_0$ and $C^{(3)}\to C^{(3)}-\mathcal{B}\wedge d\xi_0$, is
\begin{equation}
\widetilde{F}_4:=d C^{(3)}- C^{(1)}\wedge d\mathcal{B}.
\end{equation}
The Bianchi identity reads now
\begin{equation}
d\widetilde{F}_4=-dC^{(1)}\wedge d\mathcal{B},
\end{equation}
and if we regard the right hand side  of this equation as a kind of Maxwell current $\star\mathcal{J}^{Maxwell}$, we are allowed to define a Maxwell charge, by integration of the $4$-form field $\widetilde{F}_4$ on a four cycle. Another type of charge may be defined when we consider the Bianchi identity as an exterior derivative of a form, say
\begin{equation}
d(\widetilde{F}_4+C^{(1)}\wedge d\mathcal{B})=\star \mathcal{J}^{Page},
\end{equation}
and again we would define the conserved charge by integration. Comparing the two definitions, we have that
\begin{equation}
\mathcal{Q}_{D4}^{Page}=\mathcal{Q}_{D4}^{Maxwell}+\int_{\mathcal{C}^4} C^{(1)}\wedge d\mathcal{B}.
\end{equation}
One important feature of these charges is that the Maxwell charge is not quantized, while the Page charge satisfies a quantization condition.

Considering a fixed point in the radial coordinate, the following cycle
\begin{equation}
\mathcal{C}^4=\{\theta_2,\phi_1,\phi_2,\psi_2| \psi_1=\theta_1=0 \}
\end{equation}
is particularly smooth in studying the above quantities. Let us start with the Page charge for convenience. On this cycle, the equation simplifies to
\begin{equation}
\star \mathcal{J}_{D4}= dF_4\;,
\end{equation}
and the quantized Page charge is the integral of this current in the five dimensional space whose boundary is the cycle $\mathcal{C}^4$. Therefore, using the Stokes theorem and normalizing our result, we find ($C^{(3)}|_{\mathcal{C}^4}=N_c/4 \cos\theta_2 d\phi_1\wedge d\phi_2\wedge d\psi_2$)
\begin{equation}
Q_{D4}^{Page}=-\frac{1}{8\pi^3}\int_{\mathcal{C}^4} F_4=N_c.
\end{equation}

Also, we can define the Maxwell charge in this cycle as

\begin{equation}
 Q_{D4}^{Maxwell}:= Q_{D4}-\frac{1}{4\pi^3}\int_{\mathcal{C}^4} C^{(1)}\wedge d\mathcal{B}\;,
\end{equation}
and using the RR forms that we computed, we have
\bea
-C^{(1)}\wedge dB&=&\frac{N_c(1+\gamma)}{8}\frac{\Xi}{\Sigma}\sin\theta_2d\theta_2\wedge d\phi_2\wedge d\psi_2\wedge d\phi_1\cr
&=&\frac{N_c(1+\gamma)}{8}\frac{\Xi}{\Sigma}\omega^1\wedge \omega^2\wedge \omega^3\wedge d\phi_1\;,
\eea
so
\begin{equation}
-\frac{1}{4\pi^3}\int_{\mathcal{C}^4}C^{(1)}\wedge d\mathcal{B} =\frac{\Xi}{\Sigma}(1+\gamma)N_c\;,
\end{equation}
and we see that the Maxwell charge in not quantized, but it runs along the radial direction.

\section{Field theory aspects}

The original motivation of the MNa solution was from the gauge/gravity correspondence. Since we have just found a background by T-duality of the MNa solution, 
we want to study properties of the dual gauge field theory to this background. 

\subsection{Wilson loops}

Wilson loop observables are given by (see, e.g., \cite{Nunez2009, Smilga2001, Aharony1999a, Nastase2007} for more details)
\begin{equation}
W(\mathcal{C}):=\frac{1}{N_c}\Tr P \exp \left(i\oint A_\mu dx^\mu\right),
\end{equation}
where the trace is usually taken over the fundamental representation. From the expectation value of the Wilson loop, we can compute 
the quark-antiquark ($Q\bar{Q}$) potential. Choosing a rectangular loop with sides of length $L_{Q\bar{Q}}$ in the spatial 
direction and $T$ for the time direction, with $L_{Q\bar{Q}}<<T$, as $T\rightarrow \infty$ we have the behaviour
\begin{equation}
\langle W(\mathcal{C})\rangle\sim e^{-V_{Q\bar{Q}}T}\label{wilsonloop},
\end{equation}
where $V_{Q\bar{Q}}$ is the quark-antiquark potential. 

In a confining theory, the potential behaves as
\begin{equation}
V_{Q\bar{Q}}\sim \sigma L_{Q\bar{Q}},
\end{equation}
where the constant $\sigma$ is called the \textit{QCD string tension}, so the expectation value of the Wilson loop (\ref{wilsonloop}) obeys the area law,
\begin{equation}
\langle W(\mathcal{C})\rangle\sim e^{-\sigma S},
\end{equation}
for the rectangular region considered.

In the case of $\mathcal{N}=4$ SYM, dual to $AdS_5\times S^5$, we have a holographic prescription for a supersymmetric version of the Wilson loop,
\begin{equation}
W(\mathcal{C}):=\frac{1}{N_c}Tr P \exp \left[\oint (iA_\mu \dot{x}^\mu+\theta^I X^I(x)\sqrt{\dot{x}^2})d\tau\right]\label{genwilsonloop},
\end{equation}
where $x^\mu(\tau)$ parametrizes the loop and $\theta^I$ parametrizes the sphere $S^5$ and couples to the scalars $X^I$ in ${\cal N}=4$ SYM.

The holographic prescription for the Wilson loop VEV is \cite{Maldacena1998,Rey:1998ik}, 
\begin{equation}
\langle W(\mathcal{C})\rangle\sim e^ {-S},
\end{equation}
where $S$ is the area of a string world-sheet which ends on a curve $\mathcal{C}$ at the boundary of the $AdS_5$ space. 
Since the area of the worldsheet is divergent, we need to subtract the area of the string going straight down from $U=\infty$ to $U=U_0$,
\begin{equation}
W(\mathcal{C})\sim e^ {-(S-\ell\Phi)},
\end{equation}
where $\ell$ is the perimeter of the Wilson loop contour 
$\mathcal{C}$ and $\Phi=U_\infty-U_0$. The area of the worldsheet can be computed using the Nambu-Goto action
\begin{equation}
S=\frac{1}{2\pi\alpha^ \prime}\int d\tau d\sigma (\det g_{\mu\nu} \partial_\alpha X^\mu \partial_\beta X^\nu)^{1/2},\label{ngaction}
\end{equation}
where $g_{\mu\nu}$ is the $AdS_5\times S^5$ metric. In $AdS_5\times S^5$, 
we find the behaviour $V_{Q\bar{Q}}\sim 1/L_{Q\bar{Q}}$ determined by conformal invariance, see \cite{Maldacena1998,Rey:1998ik}.

We now consider a more general background,
\begin{equation}
ds^2=-g_{tt}dt^2+g_{xx}dx^2+g_{\rho\rho}d\rho^2+g_{ij}^{int}dy^i dy^j\;,\label{metric}
\end{equation}
where we assume that the functions $(g_{tt},g_{xx},g_{\rho\rho})$ are functions of $\rho$ only. We do not fix the internal space, 
since we consider a probe string that is not excited in these directions; so the internal space has no role in the present study. 

As in AdS space, we consider a string whose ends are fixed at $x=0$ and $x=L_{Q\bar{Q}}$ at the boundary of space, $\rho\to\infty$. 
In addition, we assume that it can extend in the bulk, so that the radial coordinate of the string assumes its minimum value at $\rho_0$, and that 
by symmetry this occurs at $x=L_{Q\bar{Q}}/2$.

We choose a configuration such that 
\begin{equation}
t=\tau\quad x=x(\sigma)\quad \rho=\rho(\sigma)\;,
\end{equation}
and we compute the Nambu-Goto action (\ref{ngaction}) with relation to the metric (\ref{metric}). The induced metric on the worldsheet is $G_{\alpha\beta}=g_{\mu\nu}\partial_\alpha x^\mu\partial_\beta x^\nu$, where
\begin{equation}
G_{\tau\tau}=-g_{tt},\quad G_{\sigma\sigma}=g_{xx}\left(\frac{d x}{d\sigma}\right)^2+g_{\rho\rho}\left(\frac{d \rho}{d\sigma}\right)^2,\;\;\;
G_{\tau\sigma}=0\;,
\end{equation}
and the determinant of the worldsheet is 
\begin{align}
detG_{\alpha\beta}&=-g_{tt}g_{xx}(x^\prime)^2-g_{tt}g_{\rho\rho}(\rho^\prime)^2\nonumber\\
&\equiv -f^2(x^\prime)^2-g^2(\rho^\prime)^2\;,
\end{align}
where we have defined the functions $f^2=g_{tt} g_{xx}$ and $g^2=g_{tt} g_{\rho\rho}$. Hence we write the Nambu-Goto action as
\begin{equation}
S=\frac{T}{2\pi\alpha^\prime}\int_0^{2\pi} d\sigma\sqrt{f^2(x^\prime)^2+g^2(\rho^\prime)^2}\equiv \frac{T}{2\pi\alpha^\prime}\int_0^{2\pi} d\sigma L.
\end{equation}
Its equations of motion give
\begin{align}
\partial_\tau &\left[\frac{1}{L}(f^2 x^{\prime 2}+g^2\rho^{\prime 2})\right]=0\label{first}\\
\partial_\sigma &\left[\frac{1}{L}f^2 x^{\prime}\right]=0\label{second}\\
\partial_\sigma &\left[\frac{1}{L}g^2 \rho^{\prime}\right]=\frac{1}{L}(x^{\prime2}f f^{\prime}+\rho^{\prime 2} g g^{\prime}).\label{third}
\end{align}
The first of these equations is trivially satisfied since we assume our background time independent. The second, (\ref{second}), is satisfied 
if we assume that the term inside brackets is a constant $C_0$. That means 
\begin{equation}
\frac{1}{L}f^2 x^{\prime}=C_0\, \Rightarrow\, \frac{f^2 x^{\prime }}{C_0}=(f^2 x^{\prime 2}+g^2\rho^{\prime 2})^{1/2}\;,
\end{equation}
which implies that
\begin{equation}
\frac{d\rho}{d\sigma}=\pm \frac{dx}{d\sigma}\frac{f}{C_0 g}\sqrt{f^2-C_0^2}\equiv \pm \frac{dx}{d\sigma}W_{eff}\;,
\end{equation}
thus we write 
\begin{equation}
\frac{d\rho}{d\sigma}=\pm \frac{dx}{d\sigma}W_{eff} \Rightarrow \frac{d\rho}{dx}=\pm W_{eff}.
\end{equation}
Here we wrote $W_{eff}$ just for convenience and one can check that the third equation (\ref{third}) is satisfied once we assume that the above equation is true.

From the sort of solution we are looking for, one can show that there are two distinct regions
\begin{align}
x< L_{Q\bar{Q}}/2 &\qquad \frac{d\rho}{dx}=- W_{eff}\\
x> L_{Q\bar{Q}}/2 &\qquad \frac{d\rho}{dx}= W_{eff},
\end{align}
and we can formally integrate these equations, so that
\begin{align}
\frac{d\rho}{dx}&=- W_{eff}\, \Rightarrow\, \int_{\infty}^\rho\frac{d\rho}{W_{eff}}=-\int_0^x dx \, \Rightarrow\, x(\rho)=\int^{\infty}_\rho\frac{d\rho}{W_{eff}},\quad x< L_{Q\bar{Q}}/2\\
\frac{d\rho}{dx}&= W_{eff}\, \Rightarrow\, \int^{\infty}_\rho\frac{d\rho}{W_{eff}}=\int_x^{L_{Q\bar{Q}}} dx \, \Rightarrow\, x(\rho)=L_{Q\bar{Q}}-\int^{\infty}_\rho\frac{d\rho}{W_{eff}},\quad x> L_{Q\bar{Q}}/2 .
\end{align}

The fact that the string must be fixed at $\rho\to\infty$ and we must have $x(\rho)$ finite implies that the following condition must be satisfied
\begin{equation}
\lim_{\rho\to \infty}W_{eff}(\rho)\to\infty.
\end{equation}
Once this equation is satisfied, the string moves to smaller values of the radial coordinate down to a {\it turning point} $\rho_0$ where $\frac{d\rho}{dx}|_{\rho_0}=0 $, namely where $W_{eff}(\rho_0)=0$. We restrict ourselves to turning points $C_0=f(\rho_0)$. 

Now we can compute the quark-antiquark separation pair and its potential energy. The separation is written as
\begin{equation}
L_{Q\bar{Q}}(\rho_0)=2\int_0^{L_{Q\bar Q}/2}dx=2\int^{\infty}_{\rho_0}\frac{d\rho}{W_{eff}}.
\end{equation}
In order to compute the potential $V_{Q\bar{Q}}$ we need the Nambu-Goto action $S_{NG}/T$ which diverges, but we 
need to subtract the W-boson mass given by a string going straight down on $\rho$ at $x=$const., i.e.
\begin{equation}
M=\int_0^\pi\sqrt{g^2\rho'^2}=\int_{\rho_0}^{\infty} g(\rho)d\rho,
\end{equation}
so that the renormalized quark-antiquark potential is given by
\begin{equation}
2\pi\a'V_{Q\bar{Q}}(\rho_0)=f(\rho_0)L_{Q\bar{Q}}(\rho_0)+2\int_{\rho_0}^\infty dz \frac{g(z)}{f(z)}\sqrt{f^2(z)-f^2(\rho_0)}-2 \int_{\rho_0}^\infty g(z)dz,
\end{equation}
and one can show that
\begin{equation}
2\pi \a'\frac{dV_{Q\bar{Q}}}{dL_{Q\bar{Q}}}=f(\rho_0).
\end{equation}

We can now compute the Wilson loops for the T-dual of the MNa solution. 
In this case, the solution of the set of equations is not exactly known, but remember that in the UV limit (where we consider the cutoff  
$r\sim\Lambda$) we have the asymptotic expansion  (\ref{uvi} -- \ref{uvf}), so that
\begin{align}
f^2&=g_{tt}g_{xx}\simeq e^{2\phi_\infty}\\
g^2&=g_{tt}g_{rr}\simeq e^{2\phi_\infty}N_c c_1 e^{4\Lambda/3},
\end{align}
therefore, one may check the boundary condition to see that
\begin{equation}
\lim_{r\to\Lambda}W_{eff}\sim \frac{1}{f(r_0) e^{2\Lambda/3} N_c^{1/2} c_1^{1/2}}\sqrt{e^{2\phi_\infty}-f^2(r_0)},
\end{equation}
where we will take $r_0\sim 0$, implying $f^2(r_0)=e^{2\phi_0}$.
A similar situation occurred in \cite{Caceres2014}, where the authors found a finite value for the boundary condition $\lim_{r\to\infty} W_{eff}$ and it was argued that the QFT needs to be UV-completed. Under this condition, we can calculate the QCD 
string tension (see \cite{Nunez2009, Arias2010, Sonnenschein2000}) through
\begin{equation}
\sigma=\left.\frac{1}{2\pi\alpha^\prime}f(r_0)\right|_{IR}=\frac{1}{2\pi\alpha^\prime}e^{\phi_0}\;,
\end{equation}
and therefore
\begin{equation}
2\pi \a'\frac{dV_{Q\bar{Q}}}{dL_{Q\bar{Q}}}=f(r_0)\Rightarrow V_{Q\bar{Q}}\simeq \frac{e^{\phi_0}}{2\pi\a'}L_{Q\bar{Q}}\;,
\end{equation}
which means that this theory exhibits linear confinement.

\subsection{Gauge coupling}

We can consider now another important quantity, the gauge coupling. Consider the Dirac-Born-Infeld action for a generic probe D$p$-brane, 
wrapping an $n$-cycle $\Sigma$, with induced metric 
\begin{equation}
ds_{Dp}^2=e^{2A}\eta_{\mu\nu} dx^\mu dx^\nu+ds_{\Sigma}^2
\end{equation}
and components given by $M=\{\mu,a\}$, where $\mu=0,\dots, p-n$ are indices in the Minkowski space and $a=1,\dots, n$ 
are indices of the cycle. We also take the gauge field and the Kalb-Ramond field with non vanishing components 
$F_{\mu\nu}$ and $B_{ab}$. Therefore, the DBI action reads
\begin{equation}
\begin{split}
S_{DBI}&=-T_{Dp}\int d^{p+1}\sigma e^{-\phi}\sqrt{-\det(G_{MN}+B_{MN}+2\pi\alpha^\prime F_{MN})}\\
&=-T_{Dp}\int_{\mathcal{M}} d^{p+1-n}\vec{x}\sqrt{-\det(G_{\mu\nu}+2\pi\alpha^\prime F_{\mu\nu})} \int_{\Sigma^n} d^{n}\Sigma e^{-\phi}\sqrt{-\det(G_{ab}+B_{ab})},
\end{split}
\end{equation} 
where $\mathcal{M}$ stands for Minkowski space and $d=p+1-n$ is the dimension of the reduced field theory. 
Taking an expansion of the first integral in terms of $\alpha^\prime$, we get
\begin{equation*}
S_{DBI}=-T_{Dp}\int_{\Sigma^n} d^{n}\Sigma e^{-\phi}\sqrt{-\det(G_{ab}+B_{ab})}\int_{\mathcal{M}} d^d\vec{x} e^{dA}\left( 1+\frac{(2\pi\alpha^\prime)^2 e^{-4A}}{4}F_{\mu\nu}F^{\mu\nu}+\cdots\right)\;,
\end{equation*} 
so that we can recognize the gauge coupling as 
\begin{equation}
\frac{1}{g_{YM}^2}=T_{Dp}(2\pi\alpha^\prime)^2\int_{\Sigma^n} d^{n}\Sigma e^{-(4-d)\phi-A}\sqrt{-\det(G_{ab}+B_{ab})}.
\end{equation}

Consider first the MNa solution. In this case, the induced metric on the brane is 
\begin{equation}
ds^2_{\it ind}=e^\varphi \left[dx_{1,2}^2+\frac{N_c}{4}\left(e^{2h}+\frac{e^{2g}}{4}(1-w)^2\right)(\sigma^i)^2\right],
\end{equation} 
therefore neglecting numerical factors, the coupling constant is given by
\begin{equation}
\frac{1}{g_{YM}^2}\sim \left(e^{2h}+\frac{e^{2g}}{4}(1-w)^2\right)^{3/2}\;,
\end{equation}
and using the asymptotic expansions for these functions, we see that in the IR limit, the coupling constant diverges $g_{YM}\to \infty$, whilst in the UV limit the coupling constant vanishes $g_{YM}\to 0$, and this fact is consistent with confinement and asymptotic freedom respectively, as it should be.

Now, we need to consider the case for the T-dual solution of the MNa. As we know, we need to consider first the case of the D$4$-brane wrapping a $2$-cycle defined by 
\begin{equation}
\mathcal{C}^2=\lbrace\psi_1=\psi_2\equiv \psi; \theta_1=\theta_2\equiv \theta\rbrace\;,
\end{equation}
with $\phi_1$ and $\phi_2$ fixed. Therefore, the induced metric is given by 
\begin{equation}
d\tilde{s}_{\it ind}^2=e^{2\tilde{\varphi}}\widetilde{\Sigma} dx_{1,2}^2+(\Sigma+2\Xi+\Omega)d\theta^2+\left(\Sigma\sin^2\theta+2\Xi\sin^2\theta+\Omega-\frac{\Xi^2}{\Sigma}\cos^2\theta\right)d\psi^2\;,
\end{equation}
and since the Kalb-Ramond field vanishes in this cycle, we can compute the determinant of the induced metric easily. In fact, up to numerical factors the gauge coupling is
\begin{equation}
\frac{1}{g_{YM}^2}\sim \sqrt{\widetilde{\Sigma}}e^{-\phi}
(\Sigma+2\Xi+\Omega)^{1/2}\int_{S^2}\left(\Sigma\sin^2\theta+2\Xi\sin^2\theta+\Omega-\frac{\Xi^2}{\Sigma}\cos^2\theta\right)^{1/2}\;,
\end{equation}
and the bracket inside the integral can be written as
\be
\Omega-\frac{\Xi^2}{\Sigma}+\sin^2\theta\left(\Sigma+2\Xi+\frac{\Xi^2}{\Sigma}\right)\;,
\ee
whereas 
\begin{equation}
\Sigma+2\Xi+\Omega=\Sigma-w\Omega.
\end{equation}
All terms, $\sqrt{\widetilde{\Sigma}}e^{-\phi}$, $(\Sigma+2\Xi+\Omega)^{1/2}$, $\Omega-\Xi^2/\Sigma$ and $\Sigma+2\Xi+\Xi^2/\Sigma$, go to infinity at 
$r\rightarrow\infty$, so $1/g^2_{YM}\rightarrow 0$. At $r\rightarrow 0$, $\sqrt{\widetilde{\Sigma}}e^{-\phi}$ goes to a constant, whereas $\Sigma-w\Omega$, 
$\Omega-\Xi^2/\Sigma$ and $\Sigma+2\Xi+\Xi^2/\Sigma$ go to 0 as $r^2$, so $1/g^2_{YM}\rightarrow 0$. Therefore we again have confinement ($g^2_{YM}\rightarrow
\infty$ as $r\rightarrow 0$) and asymptotic freedom ($g^2_{YM}\rightarrow 0$ as $r\rightarrow \infty$).

\subsection{Non-locality and entanglement entropy}

Another useful quantity is the entanglement entropy (EE), which can be defined as the von Neumann entropy for a reduced system, in a sense that we will 
explain below.

Consider a quantum mechanical system (we closely follow the formalisms presented in \cite{Ryu2006, Nishioka2009, Klebanov2008}), described by a pure ground state $|\Psi\rangle.$ The density matrix is 
\begin{equation}
\rho_{tot}=|\Psi\rangle\langle\Psi|
\end{equation}
and it is easy to see that the von Neumann entropy $$S_{tot}:=-\Tr(\rho_{tot}\ln \rho_{tot})$$ vanishes. By an imaginary process, we can divide the total systems into two subsystems A and B, so that, the total Hilbert space is given by the direct product of the corresponding subsystems Hilbert spaces, that is $\mathcal{H}=\mathcal{H}_A\otimes \mathcal{H}_B$.

We may think of the EE as the entropy felt by an observer who has access only to the subsystem A. Such observer will think that the system is described by the reduced density matrix
\begin{equation}
\rho_{A}=\Tr_B \rho_{tot},
\end{equation}
where we have smeared out the information of the subsystem B, by taking the trace over the Hilbert space $\mathcal{H}_B$. Then
the entanglement entropy is defined as the von Neumann entropy for the reduced system A, that is
$$S_{A}:=-\Tr(\rho_{A}\ln \rho_{A}).$$

In a $(d+1)-$dimensional QFT, it has been proved that the entanglement entropy diverges, but after introducing an ultraviolet cut-off $\varepsilon$, the divergence behaves as
\begin{equation}
S_A\propto \frac{Area(\partial A)}{\varepsilon^{d-1}}+subleading\ terms\;,
\end{equation}
since the entanglement between the subsystems A and B is more severe at the boundary $\partial A$. 

For our purposes, we can take the QFT defined on $\mathbb{R}^{d+1}$ with the following intervals \footnote{At fixed time, $t=t_0$.} \cite{Klebanov2008, Caceres2014, Kol2014}, 
\begin{align}
A&=\mathbb{R}^{d-1}\times \mathcal{I}_{\ell}\nonumber\\
B&=\mathbb{R}^{d-1}\times \mathbb{R}\backslash \mathcal{I}_{\ell}
\end{align}
where $\mathcal{I}_{\ell}$ is a line segment of length $\ell$. In such a case, the entanglement entropy is 
\begin{equation}
S_A\propto \frac{Vol(\mathbb{R}^{d-1})}{\varepsilon^{d-1}},
\end{equation}
where $Vol(\mathbb{R}^{d-1})$ is the volume of the space $\mathbb{R}^{d-1}$, since the boundary of the $d-$dimensional region A are two copies of the space $\mathbb{R}^{d-1}$ with separation $\ell$.

The computation of the EE in a QFT is not an easy task for an arbitrary region $A$, even if we consider a free theory. If we consider a theory 
with a gravity dual, we can compute the EE using the holographic prescription of \cite{Ryu2006}. In a large $N_c$ $(d+1)-$dimensional CFT, we 
find the minimal area of the $d-$dimensional surface $\gamma$ in the $(d+2)$-dimensional AdS space at $t=t_0$, whose boundary of $\gamma$ 
coincides with the boundary of the region $A$, that is $\partial\gamma=\partial A$.

The holographic entanglement entropy is given by the area of this surface 
\begin{equation}
S_A=\frac{1}{4G^{(d+2)}_N} \int_{\gamma} d^d\sigma \sqrt{G^{(d)}_{ind}}\;,
\end{equation}
where the $G^{(d)}_{ind}$ is the induced string frame metric on the surface $\gamma$. Considering a ten-dimensional metric, 
we need to take into account the fact that in non-conformal theories the dilaton and the volume of the internal space are not constant, 
therefore a natural generalization is the prescription
\begin{equation}
S_A=\frac{1}{4G^{(10)}_N} \int_{\gamma} d^8\sigma e^{-2\phi} \sqrt{G^{(8)}_{ind}}\ .\label{presc}
\end{equation}
The entropy is obtained by minimizing the action (\ref{presc}) above, over all surfaces that approach the boundary of the entangling region A. Klebanov, Kutasov and Murugan found in \cite{Klebanov2008} that in a confining background there are two surfaces minimizing the action, the first one is \textit{disconnected} which consists of two cigars descending straight down to the IR cut-off $r_0$, 
separated by a distance $\ell$, and the second is a \textit{connected} surface, in which the cigars are connected by a tube with the width depending on $\ell$.

Consider a gravitational background in the string frame of the form \cite{Klebanov2008}
\begin{equation}
ds^2=\alpha(r)[\beta(r)dr^2+\eta_{\mu\nu}dx^\mu dx^\nu]+g_{ij}^{int}dy^i dy^j \label{background}
\end{equation}
where $x^\mu \; (\mu=0,1,\dots,d)$ parametrize the flat space $\mathbb{R}^{d+1}$, $r$ is the radial coordinate and $\theta^i\; (i=d+2,\dots,9)$ are internal coordinates. The volume of the internal manifold is
\begin{equation}
V_{int}=\int d^6y\sqrt{\det [g_{ij}^{int}]},\label{volume}
\end{equation}
and if we plug the background (\ref{background}), into the prescription (\ref{presc}), we get
\begin{align}
S_A &=\frac{1}{4G^{(10)}_N} \int_{\mathbb{R}^{d-1}} d^{d-1}x\int d^6y \sqrt{\det [g_{ij}^{int}]} \int_{-\ell/2}^{+\ell/2}dx e^{-2\phi}\alpha(r)^{d/2} \sqrt{1+\beta(r)(\partial_x r)^2}\nonumber\\
&=\frac{1}{4G^{(10)}_N} Vol(\mathbb{R}^{d-1}) \int_{-\ell/2}^{+\ell/2}dx\ e^{-2\phi}V_{int}\alpha(r)^{d/2} \sqrt{1+\beta(r)(\partial_x r)^2}\nonumber\\
&=\frac{1}{4G^{(10)}_N} Vol(\mathbb{R}^{d-1})\int_{-\ell/2}^{+\ell/2}dx \sqrt{H(r)} \sqrt{1+\beta(r)(\partial_x r)^2}\;,\label{action}
\end{align}
where we have denoted by $x$ the direction along which the interval $\mathcal{I}_\ell$ lies, and also we have defined the useful quantity
\begin{equation}
H=e^{-4\phi}V_{int}^2\alpha^d\ .\label{quantity}
\end{equation}
We need to find the solution for the equation of motion in the integral (\ref{action}). Since this integral does not depend explicitly on $x$, 
we argue that the ``energy'' defined with respect to it is conserved \cite{Lewkowycz2012}, that is, if we take $\mathcal{L}=\sqrt{H+H\beta (r^\prime)^2}$, then
$$
\frac{d}{dx}\left(\frac{d \mathcal{L}}{dr^\prime}r^\prime-\mathcal{L}\right)=0
$$
implies that
\begin{equation}
\frac{d}{dx}\left(\frac{H(r)}{\sqrt{H+H\beta (r^\prime)^2}}\right)=0\;,
\end{equation}
and after fixing the constant at the minimum value of the radial coordinate $r^\ast$, we have the solution 
\begin{equation}
\frac{dr}{dx}=\frac{1}{\sqrt{\beta(r)}}\left(\frac{H(r)}{H(r^\ast)}-1\right)^{1/2}\ , \label{rprime}
\end{equation}
and integrating between $r^\ast$ and infinity, we obtain 
\begin{equation}
\frac{\ell(r^\ast)}{2}=\sqrt{H(r^\ast)}\int_{r^\ast}^{r_{\infty}} dr \left(\frac{\beta(r)}{H(r)-H(r^\ast)}\right)^{1/2} \ .
\end{equation}
Finally, we insert equation (\ref{rprime}) into (\ref{action}), and we get the entropy density for the connected solution,
\begin{equation}
\frac{S_A}{ Vol(\mathbb{R}^{d-1})}=\frac{1}{2G^{(10)}_N}\int_{r^ \ast}^{r_{\infty}}dr \frac{\sqrt{\beta(r)}H(r)}{ \sqrt{H(r)-H(r^\ast)}}\;,
\end{equation}
where we write the UV cut-off $r_{\infty}$. As we already know, the entanglement entropy generally is UV divergent, but KKM found that
the difference between the EE  of the connected and disconnected solutions is finite, and is easily seen to be given by
\begin{equation}
\frac{2G_N^{(10)}}{{ Vol(\mathbb{R}^{d-1})}}\left(S^{(c)}-S^{(d)}\right)=\int_{r_\ast}^\infty dr  \frac{\sqrt{ \beta H}}{\sqrt{1-H(r_\ast)/H(r)}}-\int_{r_0}^{\infty} dr \sqrt{\beta H}\ .
\end{equation}
The EE can be used as an order parameter for the confinement/deconfinement phase transition in a confining theory. In fact, a similar phase transition 
was found by KKM in \cite{Klebanov2008}, where they showed that depending on the value of $\ell$, the relevant solutions can be either the 
connected or the disconnected solutions and the phase transition between these two solutions is a characteristic of confining theories.

Moreover, in \cite{Kol2014}, it was proved that a sufficient condition for the existence of 
phase transitions is that the length $\ell(r_\ast)$ has an upper bound, and the non-existence of this maximum correlates with the absence of the phase transition.

We note that the quantity (\ref{quantity}) is related to the warp factor we get after a dimensional reduction on the $(8-d)$-dimensional 
compact manifold. 

In our particular case, the metric (\ref{dualmetric}) can be written as
\begin{equation}
d\tilde{s}^2_{st}=e^{2\tilde{\varphi}}\widetilde{\Sigma}dx^2_{1,2}+e^{2\tilde{\varphi}}\widetilde{\Sigma}(e^{2g}N_c)dr^ 2+\tilde{g}_{ij}^{int}dy^idy^j,
\end{equation}
so that we can compute the volume of the internal manifold (\ref{volume}) and the warp factor (\ref{quantity}) and find
$H=\widetilde{\Sigma}\sqrt{\tilde g_{int}}$, as well as $\b=e^{2g}N_c$. 

Using the metrics presented in the section (\ref{nsnssec1})  we can find
\begin{align}
l(r_\ast)&=2\sqrt{N_c H(r_\ast)}\int_{r_\ast}^\infty dr\frac{e^g}{\sqrt{H(r)-H(r_\ast)}}\label{lrast}\\
\frac{2G_N^{(10)}}{Vol(\mathbb{R}^{d-1})}\left(S^{(conn)}-S^{(disconn)}\right)&\sim N_c\int_{r_\ast}^\infty dr e^g \sqrt{H}\left(\frac{1}{\sqrt{1-H(r_\ast)/H(r)}}-1\right)\cr
&-N_c\int_{r_0}^{r_\ast} dr e^g \sqrt{H}.\label{sconndisconn}
\end{align}

One could in principle compute the volume of the internal manifold (\ref{volume}), but this gives us a very complicated equation. 
We then would need to do the following: firstly, evaluate the determinant of the internal metric and then solve the integral.

But we cannot solve analytically the integral, since we just have asymptotic solutions. We can nevertheless find the behavior of $V_{int}$.

The asymptotic behavior of the determinant is important, so we need to know - at least qualitatively - its expression. In fact, the metric of the internal manifold is of the form
\begin{equation}
[\tilde{g}^{int}]=\begin{pmatrix}
\tilde{g}_{\tilde{\theta}\tilde{\theta}} & \tilde{g}_{\tilde{\theta}\tilde{\phi}} & \tilde{g}_{\tilde{\theta}\tilde{\psi}} & \tilde{g}_{\tilde{\theta}\theta} & \tilde{g}_{\tilde{\theta}\phi} & \tilde{g}_{\tilde{\theta}\psi} \\ 
\tilde{g}_{\tilde{\phi}\tilde{\theta}} & \tilde{g}_{\tilde{\phi}\tilde{\phi}} & \tilde{g}_{\tilde{\phi}\tilde{\psi}} & \tilde{g}_{\tilde{\phi}\theta} & \tilde{g}_{\tilde{\phi}\phi} & \tilde{g}_{\tilde{\phi}\psi} \\
\tilde{g}_{\tilde{\psi}\tilde{\theta}} & \tilde{g}_{\tilde{\psi}\tilde{\phi}} & \tilde{g}_{\tilde{\psi}\tilde{\psi}} & \tilde{g}_{\tilde{\psi}\theta} & \tilde{g}_{\tilde{\psi}\phi} & \tilde{g}_{\tilde{\psi}\psi} \\
\tilde{g}_{\theta\tilde{\theta}} & \tilde{g}_{\theta\tilde{\phi}} & \tilde{g}_{\theta\tilde{\psi}} & \tilde{g}_{\theta\theta} & \tilde{g}_{\theta\phi} & \tilde{g}_{\theta\psi} \\
\tilde{g}_{\phi\tilde{\theta}} & \tilde{g}_{\phi\tilde{\phi}} & \tilde{g}_{\phi\tilde{\psi}} & \tilde{g}_{\phi\theta} & \tilde{g}_{\phi\phi} & \tilde{g}_{\phi\psi} \\
\tilde{g}_{\psi\tilde{\theta}} & \tilde{g}_{\psi\tilde{\phi}} & \tilde{g}_{\psi\tilde{\psi}} & \tilde{g}_{\psi\theta} & \tilde{g}_{\psi\phi} & \tilde{g}_{\psi\psi} 
\end{pmatrix}\;, 
\end{equation}
where the non-vanishing components are 
$$
\begin{array}{|l|l|l|}
\hline
 \tilde{g}_{\tilde{\theta}\tilde{\theta}}=\Sigma  &  \tilde{g}_{\tilde{\phi}\tilde{\phi}}=\Sigma^{-1}  & \tilde{g}_{\tilde{\psi}\tilde{\psi}}=\Sigma\sin^2\theta_1 \\ \hline \hline
\multicolumn{3}{|l|}{\tilde{g}_{\theta\theta}=\Omega-\frac{1}{\Sigma}\Xi^2\sin^2(\psi_1-\psi_2)\sin^2\theta_1} \\ 
\hline
\multicolumn{3}{|l|}{\tilde{g}_{\theta\phi}=-\frac{1}{\Sigma}\Xi^2\sin(\psi_1-\psi_2)\sin\theta_1[\sin\theta_1\sin\theta_2\cos(\psi_1-\psi_2)+\cos\theta_1\cos\theta_2] }\\ 
\hline
\multicolumn{3}{|l|}{\tilde{g}_{\theta\psi}=-\frac{1}{\Sigma}\Xi^2\sin(\psi_1-\psi_2)\sin\theta_1\cos\theta_1}\\ 
\hline
\multicolumn{3}{|l|}{\tilde{g}_{\phi\phi}=\Omega-\frac{1}{\Sigma}\Xi^2[\sin\theta_1\sin\theta_2\cos(\psi_1-\psi_2)+\cos\theta_1\cos\theta_2]^2}\\ 
\hline
\multicolumn{3}{|l|}{\tilde{g}_{\phi\psi}=\Omega\cos\theta_2-\frac{1}{\Sigma}\Xi^2\cos\theta_1[\sin\theta_1\sin\theta_2\cos(\psi_1-\psi_2)+\cos\theta_1\cos\theta_2]}\\ 
\hline
\multicolumn{3}{|l|}{\tilde{g}_{\psi\psi}=\Omega-\frac{1}{\Sigma}\Xi^2\cos^2\theta_1}\\ 
\hline \hline
\tilde{g}_{\tilde{\theta}\theta}=\Xi \cos(\psi_1-\psi_2) & \tilde{g}_{\tilde{\theta}\phi}=-\Xi \sin(\psi_1-\psi_2)\sin\theta_2 & \tilde{g}_{\tilde{\psi}\theta}=-\Xi\sin\theta_1 \cos\theta_1\sin(\psi_1-\psi_2) \\ 
\hline
\multicolumn{3}{|l|}{\tilde{g}_{\tilde{\psi}\phi}=\Xi[\cos\theta_2 \sin^ 2\theta_1-\cos\theta_1\sin\theta_1\sin\theta_2\cos(\psi_1-\psi_2)]}\\ 
\hline
\multicolumn{3}{|l|}{\tilde{g}_{\tilde{\psi}\psi}=\Xi\sin^ 2\theta_1}\\ 
\hline
\end{array}
$$

The determinant of this matrix is really laborious to calculate. However, the volume element acts just in the angular directions, 
$0\leq \theta_a \leq \pi$, \ $0\leq \phi_a < 2\pi$, \ $0\leq \psi_a < 4\pi$. So, we can ignore the expression of the angular directions, 
since it only gives us numerical factors, which in the asymptotic limit are not important at all. We are mainly interested in the radial direction.

In the UV limit $r\rightarrow \infty$, the determinant is a function of the form 
\begin{equation}
\det[\tilde{g}^{int}]\sim e^{16r/3}A+{\rm subleading}\;,
\end{equation}
where $A$ is a function of the angular directions only, so $V_{int}$ diverges at $r\rightarrow \infty$. 
We also then find $H\sim e^{16r/3}$ and $e^g\sim e^{2r/3}$, so $l(r^\ast)$ in (\ref{lrast})
and $S^{(conn)}-S^{(disconn)}$ in (\ref{sconndisconn}) are actually convergent at $r\rightarrow \infty$.

We also obtain that, modulo possible cancellations, $\det[\tilde g_{int}]$ is finite at $r\rightarrow 0$, therefore both $H$ and $\b$ remain finite at 
$r\rightarrow 0$. 

Then from (\ref{lrast}), as $r^\ast \rightarrow 0$, 
$l(r^\ast)$ goes to a constant, whereas at $r^\ast\rightarrow \infty$,
\be
l(r^\ast)\sim e^{8r_*/3}\int_{r^\ast}^\infty dr\frac{e^{2r/3}}{\sqrt{e^{16r/3}-e^{16r^\ast/3}}}=(\tilde r^\ast)^4\int_{\tilde r^\ast}^\infty\frac{d\tilde r}{
\sqrt{\tilde r^8-(\tilde r^\ast)^8}}=\tilde r^\ast \int_1^\infty \frac{dz}{\sqrt{z^8-1}}\;,
\ee
where $\tilde r=e^{2r/3}$ and $z=\tilde r/\tilde r^\ast$,
so $l(r^\ast)$ goes to infinity. This behaviour ($l(r^\ast)$ increasing to infinity) already suggests there is no phase transition. Indeed,
as was pointed in \cite{Caceres2014, Kol2014}, the 
absence of a maximum value for $l(r_\ast)$ suggests the absence of a first order phase transition in the entanglement entropy
(in the cases with phase 
transition in the entanglement entropy, we have a maximum for $l(r^\ast)$: $l$ increases to a maximum, then decreases with $r^\ast$).
To verify this, we check the sign of $S^{(conn)}-S^{(disconn)}$ at zero and infinity. 
At $r^\ast\rightarrow 0$, 
\be
\Delta S|_{r^\ast\rightarrow 0}\sim \int_{r^\ast\rightarrow 0}^\infty dr e^g(r) \sqrt{H(r)}\left(\frac{1}{\sqrt{1-H(r^\ast)/H(r)}}-1\right)>0\;,
\ee
since the integrand is positive. At $r^\ast \rightarrow \infty$,
\bea
\Delta S|_{r^\ast\rightarrow \infty}&\sim& \int_{r^\ast\rightarrow\infty}^\infty dr e^g(r)\sqrt{H(r)}\left(\frac{1}{\sqrt{1-H(r^\ast)/H(r)}}-1\right)
-\int_0^{r^\ast\rightarrow\infty} dr e^g(r)H(r)\cr
&\sim & \int_{r^\ast\rightarrow\infty}^\infty dr e^{\frac{10r}{3}}\left(\frac{1}{\sqrt{1-e^{\frac{8(r^\ast -r)}{3}}}}-1\right)-\int_0^{r^\ast\rightarrow\infty}
dr e^{\frac{10r}{3}}\cr
&=&\frac{3}{2}(\tilde r^\ast)^5\left[\int_1^\infty dz \; z^4\left(\frac{1}{\sqrt{1-z^{-8}}}-1\right)-\int_0^1 dz \; z^4\right]\cr
&=&\frac{3}{2}(\tilde r^\ast)^5\frac{\sqrt{\pi}\Gamma\left(\frac{3}{8}\right)}{40\Gamma\left(\frac{7}{8}\right)}\rightarrow +\infty\;,
\eea
so is not only positive, but goes to infinity. If nothing strange happens in between (at finite $r^\ast$), 
it means that the disconnected solution has always the lower entropy, implying that there is no phase transition.
It is worth mentioning 
that this behavior is consistent with \cite{Kol2014}, where a detailed study of entanglement entropy as a probe of confinement was considered. In fact, 
they showed that the UV completion done in \cite{Gaillard2011} provides a consistent model with phase transitions.

\subsection{Domain walls}

Our configuration consists of a D4-Brane wrapping a two-cycle defined by $\mathcal{C}^2=\{\theta_1=\theta_2,\psi_1=\psi_2\}$ and for $\phi_1=$const., 
this cycle vanishes in the IR limit. 

We may think of probe D$4$ branes that wrap the cycle $\mathcal{S}^2=\{\theta_1,\psi_1\}$ at $r\to 0$ and the remaining angular directions are fixed. This configuration can act as a domain wall if it has finite tension. This is an useful observable, since even in the presence of singularities, the tension of the domain wall remains finite. Taking the cycle $S^2$, the induced metric is
\begin{equation}
d\tilde{s}^{2}_{S^2}=\frac{N_c e^{2\tilde{\varphi}}}{4}\left(e^{2h}+\frac{e^{2g}}{4}(1+w)^2 \right) dx^{2}_{1,2}+\Sigma (d\theta_1^2+\sin^2\theta_1 d\psi_1^2).
\end{equation}
The tension of the domain wall can be computed from the DBI action of the $D4$-brane 
\begin{equation}
S=-T_{D4}\int d\theta_1 d\psi_1\int d^3 x e^{-\tilde{\varphi}}\sqrt{|\tilde{g}|}\equiv -T_{eff}\int d^3 x,
\end{equation}
so that the tension in the IR,
\begin{equation}
T_{eff}=4\pi e^{-\phi/2}\left(\frac{N_c}{4}\right)^2\left(e^{2h}+\frac{e^{2g}}{4}(1+w)^2\right)^2\Sigma T_{D4}\simeq 
4\pi e^{\phi_0/2}\left(\frac{N_c}{4}\right)^3 g_0^3 T_{D4} 
\end{equation}
is finite. We can follow the formalism of \cite{Acharya2001c} (see also \cite{Caceres2014}) and add a gauge field $A_1$, with field strength $G_2=dA_1$ in the Minkowski part of the world volume of the brane, in such a way that we induce a Wess-Zumino term of the form
\begin{equation}
S_{WZ}=T_{D4}\int C^ {(1)}\wedge G_2 \wedge G_2\equiv -T_{D4}\int F_{2}\wedge G_2\wedge A_1, \label{wess-zumino}
\end{equation} 
where $C^{(1)}$ is the one-form that we found above, and $F_{2}=dC^{(1)}$ its field strength. Using the cycle $S^2$, in which the field strength is
\begin{equation}
\left. \phantom{\frac{1}{1}}F_2\right|_{S^2}=-\frac{N_c}{4}\sin\theta_1 d\theta_1\wedge d\psi_1,
\end{equation}
we can perform the integral
$$\int_{S^2} F_{2}=-2\pi N_c, $$ and we insert this integral into the Wess-Zumino action (\ref{wess-zumino}) above, so that
\begin{equation}
S=2\pi N_c T_{D4}\int G_2\wedge A_1.
\end{equation}
We see that we have induced a Chern-Simons term in the $2+1$ gauge theory, on the domain wall.

\section{Conclusions}

In this paper we have considered a T-duality along an $U(1)$ isometry of a deformation of the MNa solution in \cite{Maldacena2001}, 
such that the resulting type IIA solution consists of D$4$-branes wrapping a two-cycle. We found a solution with non-trivial  RR forms, a non-vanishing Kalb-Ramond field and a complicated metric. We analyzed Maxwell and Page charges associated to this solution.

We then studied properties of the field theory dual to the T-dual gravitational background. From a calculation of the Wilson loops, we saw that the 
dual gauge theory presents confinement. We also computed the QCD string tension and the gauge coupling of the gauge theory. 

From a calculation of the entanglement entropy, we found that the field theory does not have a phase transition, despite being a confining theory; 
this could be due to the non-locality of the theory, as suggested in \cite{Kol2014}. 
Finally, considering domain walls in the gravitational background, we generate a Chern-Simons term in the gauge theory.

\acknowledgments

The work of HN is supported in part by CNPq grant 301219/2010-9 and FAPESP grant 2013/14152-7 and the work of TA is supported by CNPq grant 140588/2012-4. TA is grateful to Prieslei Goulart for useful discussions.

\end{document}